# Influence of the dividing surface notion on the formulation of Tolman's law

M T Horsch

STFC Daresbury Laboratory, United Kingdom

American University of Iraq, Sulaimani (AUIS), Iraq

*martin.horsch@stfc.ac.uk*

**ABSTRACT**

The influence of the surface curvature $1/R$ on the surface tension $\gamma$ of small droplets at equilibrium with a surrounding vapour, or small bubbles at equilibrium with a surrounding liquid, can be expanded as $\gamma(R) = \gamma_0 - 2\delta_0/R + O(1/R^2)$, where $R = R_L$ is the Laplace radius and $\gamma_0$ is the surface tension of the planar interface, with zero curvature. According to Tolman's law, the first-order coefficient in this expansion is obtained from the planar limit $\delta_0$ of the Tolman length, i.e., the deviation $\delta = R_e - R_L$ between the equimolar radius $R_e$ and $R_L$. Here, Tolman's law is generalized such that it can be applied to any notion of the dividing surface, beside the Laplace radius, on the basis of a generalization of the Gibbs adsorption equation which consistently takes the size dependence of interfacial properties into account.

**Key words:** Nucleation theory; surface tension; nanodroplets; Tolman's law; interfacial thermodynamics.

## 1. INTRODUCTION

The surface tension of small droplets or bubbles at the nanometre length scale has a significant influence on the formation of dispersed phases by homogeneous nucleation (Mokshin and Galimzyanov, 2012). It is also crucial to understand droplet or bubble growth, decay, coalescence, and the coupling between heat and mass transfer at curved interfaces (Sumardiono and Fischer, 2007). The capillarity approximation is often employed: Thereby, the surface tension of a nanodroplet or nanobubble is assumed to be identical with the macroscopic value (Feder *et al.,* 1966), which is attained in the limit of zero curvature, i.e., infinite droplet radius, or for a planar interface, which can be measured experimentally and is known at a good precision for most fluids.

However, the capillarity approximation is not exact; the surface tension is known to depend on the size of the nanodispersed phase. An expression for the size dependence of the surface tension, Tolman's law (Tolman, 1949) is widely used. It relates the curvature dependence of the surface tension to a characteristic length scale, the Tolman length $\delta$, which is determined from the deviation between the equimolar radius and the Laplace radius. A simplified version of this law, suggested by Tolman (1949) himself, considers only a linear contribution of the curvature to the surface tension and neglects all higher-order terms. It should be noted, however, that the linearized version of Tolman's law was found to be insufficient for many typical applications, since the Tolman length, which controls the magnitude of the linear term, is extremely small for droplets (Homman *et al.,* 2014) as well as bubbles (Min and Berkowitz, 2019). Das and Binder (2011) observed that the most relevant contribution may come from the second-order term, and a contribution proportional to cubic curvature was postulated by Malijevský and Jackson (2012). For the hard-sphere fluid confined by hard walls, significant second-order, but negligible higher-order curvature contributions were found (Davidchack and Laird, 2018). For atomic nuclei, Cherevko *et al.* (2015) determined a Tolman length of the order of $\delta \approx 10^{-15}$ m.

The present work discusses how the formulation of Tolman's law relates to the boundary conditions employed to describe the state of the system, and how differences between notions (i.e., definitions) of the spherical dividing surface affect the way in which Tolman's law needs to be formulated. Considering the size dependence of the surface tension for spherical dividing surfaces by the Gibbs approach to interfacial thermodynamics, where the position of a formal two-dimensional dividing surface can be specified freely, a generalized version of Tolman's law is obtained, which can be combined with any notion of the dividing surface.

## 2. INTERFACIAL THERMODYNAMICS

*2.1 Notion of the dividing surface*

Interfacial thermodynamics following Gibbs (1878) is based on the definition of a dividing surface, a strictly two-dimensional object at which two discrete parts of the volume, containing two coexisting phases, are in contact with each other. Here, we consider spherical interfaces, with a *dispersed* phase $\alpha$ inside and a *surrounding* phase $\beta$ outside. The interface does not occupy any volume, so that the overall volume of the system can be decomposed into the contributions from these two parts, $V = V^\alpha + V^\beta$, without having to account for any excess or deviation expression. The two phases are assumed to be at thermodynamic equilibrium, and they are both assumed to be fluid phases (vapour-liquid or liquid-liquid equilibrium).

Since in reality, at the molecular length scale, the boundary region between two fluid phases exhibits a continuous transition from one side to the other, multiple conventions can be applied to determine where exactly the dividing surface is located. Accordingly, the volumes ascribed to each of the two phases (and many other properties) are not uniquely determined by the thermodynamic state of the system. For one and the same system at a given thermodynamic state, they depend on the choice of the dividing surface notion $v$, which is a modelling decision to denote this, such quantities will be labelled with an index $v$. Due to volume balance, as above,

$$V = (V^\alpha)_v + (V^\beta)_v. \tag{1}$$

Overall properties, such as the total volume of the system $V$, are unaffected by the choice of $v$ and therefore do not carry $v$ as an index. For a spherical dividing surface, which is characterized by the radius $R_v$, an interfacial area $A_v = 4\pi(R_v)^2$ and a volume $(V^\alpha)_v = 4\pi(R_v)^3/3$ are ascribed to the dispersed phase $\alpha$.

*2.2 Properties of single-phase reference systems*

For a system with $n$ components and a single phase, any combination of $n+1$ intensive properties is here understood to uniquely determine all intensive properties, irrespective of the size of that system, in agreement with the *equation of state* for the *macroscopic* homogeneous bulk phase. By implication, this means that all finite-size effects are ascribed to the interface. In this way, the Gibbs approach differs from small-systems thermodynamics inspired by Hill (1964), which does account for a size dependence of intensive properties for homogeneous systems. Promising efforts have been undertaken recently to accommodate Hill's thermodynamics within molecular modelling and simulation (Strøm *et al.*, 2017), showing in particular that this does not add any generality to the Gibbs formalism in interfacial thermodynamics (Bedeaux and Kjelstrup, 2018). Here, therefore, the Gibbs approach is followed by construction, and no deviation from the macroscopic equation of state is assumed to occur for the equations of state applied to the homogeneous single-phase reference systems.

Following a common notational convention (Rowlinson and Widom, 1982), vectors such as **μ** contain properties associated with the $n$ components, e.g., the chemical potentials $\mu_1, \ldots, \mu_n$. Coexisting phases at thermodynamic equilibrium have the same temperature $T$ and chemical potentials **μ**; these are $n+1$ values, which are all intensive, by which the intensive properties of a reference system for the phase $\alpha$ uniquely determine all intensive properties of the reference system for the phase $\beta$, and vice versa. This implies that the intensive properties of the homogeneous reference systems representing $\alpha$ and $\beta$ are not influenced by the choice of notion for the dividing surface. However, the extensive properties of the reference systems do depend on $v$, which controls the volumes ascribed to $\alpha$ and $\beta$; for any intensive property of one of the phases, e.g., the surrounding phase $\beta$, given by $x^\beta$ (n.b., not influenced by the notion $v$), the corresponding extensive property is

$$(X^\beta)_v = (N^\beta)_v x^\beta = \rho^\beta x^\beta (V^\beta)_v, \tag{2}$$

where $\rho^\beta = 1/v^\beta$ is the density of the phase $\beta$. Therefore, the extensive property $(X^\beta)_v$ needs to carry $v$ as a subscript.

*2.3 Properties ascribed to the interface and to nucleus formation*

*Interfacial excess quantities* (e.g., the interfacial excess entropy) are determined by comparing a property of the actual two-phase system (e.g., its entropy) with the value obtained by summation over two single-phase reference systems, without an interface, at the same temperature and chemical potentials. Generally, for any extensive property *X*, the corresponding interfacial excess quantity is defined by

$$(X^E)_v = X - (X^\alpha)_v - (X^\beta)_v. \tag{3}$$

While the value of $(X^E)_v$ generally depends on *v*, the notion cannot have any influence on quantities defined by

$$\Delta X^* = X - \rho^\beta x^\beta V, \tag{4}$$

i.e., by comparing an extensive property *X* of the actual two-phase system to a state where the dispersed phase is absent and the surrounding phase *β* (at constant intensive properties) occupies the whole volume *V* of the system (Rehner and Gross, 2018). These quantities, denoted by *ΔX\** for any extensive property *X*, characterize *nucleus formation* (i.e., dispersed-phase formation) at invariant **μ**, *V*, and *T*. By combining these definitions with Eq. (1), it can be seen that nucleus formation and interfacial excess quantities are related by

$$\Delta X^* = (X^E)_v + (\rho^\alpha x^\alpha - \rho^\beta x^\beta)(V^\alpha)_v. \tag{5}$$

For the grand potential *Ω*, the change due to critical nucleus formation *ΔΩ\** represents the free energy barrier that needs to be overcome for homogeneous nucleation; it controls the nucleation rate $J \sim \exp(-\Delta\Omega^*/kT)$.

## 3. SURFACE TENSION, ADSORPTION, AND TOLMAN'S LAW

### 3.1  Definitions of the surface tension

The surface tension can be defined as an excess quantity (per surface area)

$$\sigma_v = (\Omega^E)_v / A_v, \tag{6}$$

in terms of the grand potential, cf. Eq. (3), e.g., following Mu *et al.* (2018). The value obtained from this definition will be referred to as the *absolute surface tension*, using the symbol *σ* (with a *v* index, since the excess grand potential depends on the selected dividing surface).

Another approach to defining the surface tension is based on total differentials, e.g.,

$$d\Omega = -p^\alpha d(V^\alpha)_v - p^\beta d(V^\beta)_v - S\,dT - \mathbf{N}\,d\boldsymbol{\mu} + \gamma_v\,dA_v, \tag{7}$$

where *p* represents the pressure, *S* the entropy, and **N** the amount of substance (for all components). Following Rowlinson and Widom (1982), this can be transformed to

$$d(\Omega^E)_v = -(S^E)_v\,dT - (\mathbf{N}^E)_v\,d\boldsymbol{\mu} + \gamma_v\,dA_v. \tag{8}$$

The quantity $\gamma_v$ from Eqs. (7) and (8) is called the *differential surface tension*. The two definitions given above are in agreement ($\sigma_v = \gamma_v$) if the dividing surface is positioned such that the Laplace equation holds ($R_v = R_L$).

### 3.2  Gibbs adsorption equation

Tolman (1949) relied on the Gibbs adsorption equation to deduce Tolman's law. To reexamine Tolman's law within the formalism and notation established above, is is therefore necessary to determine how the Gibbs adsorption equation can be generalized such that it can be applied to any notion of the dividing surface. In a homogeneous system, $\Omega = -pV$, and hence, for the heterogeneous system consisting of the phases *α* and *β* plus the interface which contributes $(\Omega^E)_v$, the grand potential is

$$\Omega = -p^\alpha(V^\alpha)_v - p^\beta(V^\beta)_v + (\Omega^E)_v. \tag{9}$$

The pressures $p^\alpha$ and $p^\beta$ and the total grand potential *Ω* do not depend on the notion of the dividing surface, but the contributions ascribed to the phases and the interface do. By differentiation, with Eq. (6), this gives

$$d\Omega = -p^\alpha d(V^\alpha)_v - p^\beta d(V^\beta)_v - (V^\alpha)_v\,dp^\alpha - (V^\beta)_v\,dp^\beta + \sigma_v\,dA_v + A_v\,d\sigma_v. \tag{10}$$

Moreover, the Gibbs-Duhem equation can be applied to the two homogeneous reference systems that represent the dispersed phase and the surrounding phase,

$$(V^\alpha)_v\,dp^\alpha - (S^\alpha)_v\,dT - (\mathbf{N}^\alpha)_v\,d\boldsymbol{\mu} = 0, \tag{11}$$

$$(V^\beta)_v \, dp^\beta - (S^\beta)_v \, dT - (\mathbf{N}^\beta)_v \, \mathbf{d\mu} = 0. \tag{12}$$

If the three equations above are subtracted from Eq. (7), this simplifies to

$$-(S^E)_v \, dT - (\mathbf{N}^E)_v \, \mathbf{d\mu} + (\gamma_v - \sigma_v) \, dA_v - A_v \, d\sigma_v = 0, \tag{13}$$

which divided by the surface area $A_v$ becomes the adsorption equation

$$d\sigma_v = -\zeta_v \, dT - \mathbf{\Gamma}_v \, \mathbf{d\mu} + (\gamma_v - \sigma_v) \, d\ln A_v. \tag{14}$$

Therein, $\zeta_v = (S^E)_v / A_v$ is the (specific) surface entropy, and $\mathbf{\Gamma}_v = (\mathbf{N}^E)_v / A_v$ is the adsorption. Whenever $\gamma_v$ and $\sigma_v$ are equal, such as for the Laplace radius, the well-known isothermal expression for the Gibbs adsorption equation is obtained, $(d\sigma_v)_T = -\mathbf{\Gamma}_v \, \mathbf{d\mu}$. In general, the deviation from the standard form of the adsorption equation is given by $(\gamma_v - \sigma_v) \, d\ln A_v$, the third term in Eq. (14).

*3.3    Tolman's law*

The generalized Gibbs adsorption equation is now applied to the special case of a transition where only the chemical potential of a single component $\mu_i$ is varied, while $T$ and all $\mu_{j\neq i}$ are constant; n.b., for a single-component system, this corresponds to the case considered by Tolman (1949). Noting that $d\ln A_v = 2 \, d\ln R_v$ while (Tolman, 1949)

$$d\mu_i = dp^\alpha / \rho_i^\alpha = dp^\beta / \rho_i^\beta = d\Delta p / \Delta \rho_i, \tag{15}$$

where $\Delta p = p^\alpha - p^\beta$ and $\Delta \rho_i = \rho_i^\alpha - \rho_i^\beta$, the adsorption equation simplifies to

$$d\sigma_v + \Gamma_{i,v} (\Delta \rho_i)^{-1} \, d\Delta p = \varphi_v \, dR_v \tag{16}$$

with $\varphi_v = 2(\gamma_v - \sigma_v)/R_v$ for a transition at constant $\mu_{j\neq i}$ and $T$. Using the notation $\Lambda_{i,v} = \Gamma_{i,v} (R_v \Delta \rho_i)^{-1}$ for Tolman's dimensionless adsorption expression, this becomes

$$d\sigma_v / R_v - \varphi_v \, d\ln R_v + \Lambda_{i,v} \, d\Delta p = 0. \tag{17}$$

Following Tolman (1949),

$$\Lambda_{i,v} = \delta_{i,v} / R_v + (\delta_{i,v} / R_v)^2 + (\delta_{i,v} / R_v)^3 / 3 \tag{18}$$

is obtained by determining the adsorption from a spherical density profile. The Tolman length

$$\delta_{i,v} = R_{e,i} - R_v \tag{19}$$

is defined with respect to the equimolar radius $R_e$ in the case of a single component, following Tolman (1949); in the case of a mixture, $R_{e,i}$ refers to the dividing surface $v = v_{e,i}$ at which the adsorption of the component $i$ is zero, $\Gamma_{i,v} = 0$. By Eq. (17), a generalized version of Tolman's law is given, which can be applied to any dividing surface and down to the smallest length scale. It should be noted that if $\varphi_v = 0$ (i.e., $\sigma_v = \gamma_v$), the standard form of Tolman's law holds exactly.

## 4. CONCLUSIONS

Care must be taken when concepts from macroscopic equilibrium thermodynamics are applied to small systems (Hill, 1964). This also holds for relations which are well established, such as the Gibbs adsorption equation. It was discussed here how a deviation between absolute and differential definitions of the surface tension affects this equation. A generalized Gibbs adsorption equation was derived which takes these effects consistently into account. On this basis, Tolman's law was generalized to arbitrary notions of the dividing surface. However, it is known from previous work that Tolman's law is inadequate empirically (Wilhelmsen *et al.*, 2015). Since recent findings indicate that $\delta$ is very small or zero, and that the contribution to $\gamma$ which is proportional to $1/R$ is very small (and could be entirely absent), such approaches will probably fail to capture $\gamma(R)$ correctly. The challenge of developing rigorous methods to quantify the higher-order curvature contributions to the surface tension of small bubbles and droplets therefore remains a significant area of investigation.

**Acknowledgements:** The present work was facilitated by valuable advice from Joachim Groß and Philipp Rehner. It is based on discussions with Stefan Becker, Felix Diewald, Hans Hasse, Michaela Heier, George Jackson, Ralf Müller, Jayant Kumar Singh, and Jadran Vrabec.